\documentclass[8pt,a4paper,twocolumn,english,amssymb,nofootinbib,superscriptaddress]{revtex4-2}
\usepackage{lmodern}

\usepackage[T1]{fontenc}
\usepackage[latin9]{inputenc}
\usepackage{color}
\usepackage{babel}
\usepackage{amsmath}
\usepackage{amssymb}
\usepackage{esint}
\usepackage[unicode=true,pdfusetitle,
 bookmarks=true,bookmarksnumbered=false,bookmarksopen=false,
 breaklinks=false,pdfborder={0 0 1},backref=false,colorlinks=false]
 {hyperref}

\makeatletter

\pdfpageheight\paperheight
\pdfpagewidth\paperwidth

\@ifundefined{textcolor}{}
{%
 \definecolor{BLACK}{gray}{0}
 \definecolor{WHITE}{gray}{1}
 \definecolor{RED}{rgb}{1,0,0}
 \definecolor{GREEN}{rgb}{0,1,0}
 \definecolor{BLUE}{rgb}{0,0,1}
 \definecolor{CYAN}{cmyk}{1,0,0,0}
 \definecolor{MAGENTA}{cmyk}{0,1,0,0}
 \definecolor{YELLOW}{cmyk}{0,0,1,0}
}

\usepackage{babel}
\@ifundefined{definecolor}{\usepackage{color}}{}
\@ifundefined{definecolor}{\usepackage{color}}{}
\usepackage{babel}

\@ifundefined{definecolor}{\usepackage{color}}{}
\usepackage{babel}

\@ifundefined{definecolor}{\usepackage{color}}{}
\usepackage{latexsym}\usepackage{bm}

\makeatother

\begin{document}

\title{Minimal Einstein-Aether Theory}

\author{Metin G\"{u}rses}
\email{gurses@fen.bilkent.edu.tr}

\affiliation{{\small{}{}Department of Mathematics, Faculty of Sciences}\\
 {\small{}{}Bilkent University, 06800 Ankara, Turkey}}
\author{\c{C}etin \c{S}ent{\" u}rk}
\email{csenturk@thk.edu.tr}

\affiliation{Department of Aeronautical Engineering,\\
 University of Turkish Aeronautical Association, 06790 Ankara, Turkey}
\author{Bayram Tekin}
\email{btekin@metu.edu.tr}

\affiliation{Department of Physics, Middle East Technical University, 06800 Ankara,
Turkey}

\thispagestyle{empty}

\begin{abstract}
\noindent We show that there is a phenomenologically and theoretically consistent limit of the generic Einstein-Aether theory in which the Einstein-Aether field equations reduce to Einstein field equations with a perfect fluid distribution sourced by the aether field. This limit is obtained by taking three of the coupling constants of the theory to be zero but keeping the expansion coupling constant to be nonzero. We then consider the further reduction of this limited version of Einstein-Aether theory by taking the expansion of the aether field to be constant (possibly zero), and thereby we introduce the Minimal Einstein-Aether theory that supports the Einstein metrics as solutions with a reduced cosmological constant. The square of the expansion of the unit-timelike aether field shifts the bare cosmological constant and thus provides, via local Lorentz symmetry breaking inherent in the Einstein-Aether theories, a novel mechanism for reconciling the observed, small cosmological constant (or dark energy) with the large theoretical prediction coming from quantum field theories. The crucial point here is that minimal Einstein-Aether theory does not modify the well-tested aspects of General Relativity such as solar system tests and black hole physics including gravitational waves.

\end{abstract}
\maketitle

\section{Introduction}

One of the most difficult problems in physics today is to understand how the predicted cosmological constant coming from the zero point fluctuations of the quantum fields can differ from the observed dark energy by 50-120 orders of magnitude, a huge gap that depends on the cut-off used in the regularization of the quantum field theory. This cosmological constant problem arises within the context of General Relativity coupled minimally with the Standard Model fields.
The difficulty of the problem is related to the fact that due to the successes of both theories, there is very little room for modification as the most obvious modifications are with the gravity sector and often lead to drastic changes in General Relativity. Here we will argue that there is a modified gravity theory that allows Einstein spacetimes as solutions, and the effective cosmological constant can be reduced as a result of local symmetry breaking induced by a non-zero vacuum value of a vector field. The theory is a sub-class of the Einstein-Aether theory which we shall describe first.

The so-called Einstein-Aether theory  \cite{jm,jac} is a vector-tensor theory of gravity that breaks the local Lorentz symmetry of spacetime due to the
presence of a timelike vector field \cite{matt,liber}. The nowhere-vanishing, unit-norm vector field--called the ``aether''--defines a preferred direction at
each point of spacetime and dynamically couples to the metric tensor to preserve the background independence of the theory. Over the years, Einsten-Aether Theory has attracted a lot of attention and been investigated in various respects: For example, the stability issue of the aether was discussed in \cite{cdgt,dj}, time-independent spherically symmetric
solutions and black hole solutions were analyzed in \cite{ej1,ej2,gej,tm,bjs,gs,bs,dww,afjw}, generalizations and cosmological implications were studied in
\cite{cl,zfs1,bdfsz,zfs2,zzbfs}, extensions to include other fields were considered in \cite{bl,ab}, all G\"odel-type of solutions were given in \cite{gur1,gur2}, and the possibilities of spacelike and null aether field were discussed respectively in \cite{rizz,acw,ct,cpp} and in \cite{GS}.

Einsten-Aether Theory has four dimensionless parameters $c_{1}$, $c_{2}$, $c_{3}$, and $c_{4}$ that have certain bounds, discussed here in Sec. II, by comparing the theory with
the solar system and other observations. In general, these parameters are assumed to be non-vanishing. It is known that the Einstein-Aether Theory is an extension of some other well-studied gravity theories and reproduces these theories to some other gravity under certain
limits of the coupling constants. The simplest example is the trivial Einstein limit of  Einstein-Aether theory where all the coupling constants $\{c_{1}, c_{2}, c_{3}, c_{4}\}$ are taken to be zero. If the so-called ``twist'' coupling constant, $c_{w}\equiv c_{1}-c_{3}$, is taken to infinity, one gets the Ho\v{r}ava gravity limit  \cite{ted}. Recently, yet another limit of Einstein-Aether theory has been found by Franzin et al. in \cite{franz}: Taking $c_{1}=c_{3}=c_{4}=0$ and the expansion $\theta$ of the unit timelike vector field to be zero in  Einstein-Aether theory, one obtains the vacuum Einstein field equations in the absence of matter fields. This theory is dubbed the Minimal Einsten-Aether Theory.  In this restricted version of the Einsten-Aether Theory, there is no contribution to the resulting geometry from the  Einsten-Aether Theory itself. In \cite{franz}, it was concluded that the Kerr metric with a unit time-like vector having zero expansion is an exact solution of the  Einsten-Aether Theory. Here, in this work, we generalize this result by first considering the limit $c_{1}=c_{3}=c_{4}=0$ in general and then taking $\theta$ as constant (with zero being a special case). The last case defines the Minimal Einstein Theory we introduce here which is a nontrivial generalization of the recent work \cite{franz} in which $\theta=0$ specifically. More explicitly, by including a cosmological constant, we first show that there is a phenomenologically and theoretically consistent limit--a reduced version--of the generic Einstein-Aether theory obtained by taking $c_{1}=c_{3}=c_{4}=0$ which lead the Einstein-Aether field equations to take the form of the Einstein field equations with perfect fluid distribution sourced by the aether field. Then, with the additional assumption $\theta=const.$, we prove that the Einstein-Aether theory further reduces to the Einstein theory with a shifted cosmological constant. In fact, in this new limit, there is an important contribution to the cosmological constant in the theory which is proportional to the square of the constant expansion parameter of the unit timelike vector field.

Throughout the paper, we shall adopt the sign convention $(-,+,+,+)$.

\section{The Einstein-Aether Theory}

The general action of   Einstein-Aether theory, in the presence of a cosmological constant $\Lambda$ and the matter fields, is given by
\begin{equation} \label{action}
I={1 \over 16 \pi G}\, \int d^{4}x\sqrt{-g}\,{\cal L}+I_M(g_{\mu\nu},v^\mu,\phi),
\end{equation}
where $G$ is the bare gravitational constant related to Newton's constant $G_N$ by the rescaling \cite{cl}
\begin{equation}
G_N=\left(1-\frac{c_1+c_4}{2}\right)^{-1}G,
\end{equation}
and $I_M$ denotes the matter action with $\phi$ collectively representing the matter fields. The action of the gravity sector is constructed  from the following vector-tensor  Lagrangian density:
\begin{equation}
{\cal L}=R-2\Lambda-K^{\mu \nu}\,_{\alpha \beta}\, \nabla_{\mu}\,
v^{\alpha}\, \nabla_{\nu}\, v^{\beta}+\lambda\, (v^{\mu}\,
v_{\mu}+1),\label{lag}
\end{equation}
in which the vector  $v^\mu$, called the aether field, for non-zero values,  would break local Lorentz invariance.  And to ensure this breaking,
the field $\lambda$ is taken to be the Lagrange multiplier that enforces the unit-norm constraint
\begin{equation}\label{con}
v^{\mu}\, v_{\mu}=-1,
\end{equation}
which certainly breaks local Lorentz symmetry.  In some sense, this construction resembles Nambu's construction of electrodynamics in the nonlinear gauge \cite{Nambu}, but here the underlying vector theory is not generically the usual electrodynamics coupled to gravity as the $K$ tensor is judiciously chosen to be of the form
\begin{equation}
K^{\mu \nu}\,_{\alpha \beta}=c_{1}\, g^{\mu \nu} g_{\alpha
\beta}+c_{2}\, \delta^{\mu}_{\alpha}\,
\delta^{\nu}_{\beta}+c_{3}\, \delta^{\mu}_{\beta}
\delta^{\nu}_{\alpha}-c_{4}\, v^{\mu}\, v^{\nu}g_{\alpha
\beta},\label{Ktensor}
\end{equation}
where  $\{c_{1}, c_{2}, c_{3}, c_{4}\}$ are dimensionless coupling constants which certainly are phenomenologically bounded.

It pays to reparameterize the theory with the use of the usual hypersurface decomposition of the timelike aether congruence as \cite{ted}
\begin{equation}
\nabla_{\nu}v_{\mu}=\frac{1}{3}\theta h_{\mu\nu}+\sigma_{\mu\nu}+\omega_{\mu\nu}-a_\mu v_\nu,
\end{equation}
where $\theta$ is the expansion of the aether field, $h_{\mu\nu}$ is the projection tensor that projects onto the hypersurface orthogonal to $v_\mu$, $\sigma_{\mu\nu}$ is the shear,
$\omega_{\mu\nu}$ is the twist, and $a_\mu$ is the acceleration, which are all defined by
\begin{eqnarray}
&&\theta=\nabla_\mu v^\mu, \nonumber\\
&&h_{\mu\nu}=g_{\mu\nu}+v_\mu v_\nu,\nonumber\\
&&\sigma_{\mu\nu}=a_{(\mu}v_{\nu)}+\nabla_{(\mu}v_{\nu)}-\frac{1}{3}\theta h_{\mu\nu},\nonumber\\
&&\omega_{\mu\nu}=a_{[\mu}v_{\nu]}-\nabla_{[\mu}v_{\nu]},\nonumber\\
&&a_\mu=v^\nu\nabla_\nu v_\mu,
\end{eqnarray}
where we used the usual symmetrization and anti-symmetrization brackets with a factor of $1/2$. With these new variables, the Lagrangian in (\ref{lag})
reduces to
\begin{equation}
{\cal L}=R-2\Lambda-\frac{1}{3}c_\theta \theta^2-c_\sigma\sigma^2-c_\omega\omega^2+c_a a^2+\tilde{\lambda}\, (v^{\mu}\,
v_{\mu}+1),\label{lag1}
\end{equation}
where $c_{\theta}$, $c_{\sigma}$, $c_{\omega}$, and $c_{a}$ are the redefined dimensionless parameters with the definitions
\begin{eqnarray}
&&c_\theta\equiv c_1+c_3+3c_2,\nonumber\\
&&c_\sigma\equiv c_1+c_3,\nonumber\\
&&c_\omega\equiv c_1-c_3,\nonumber\\
&&c_a\equiv c_1+c_4,\label{redefc}
\end{eqnarray}
and $\tilde{\lambda}$ is again the Lagrange multiplier, but not necessarily the same as $\lambda$ in (\ref{lag}).

We mentioned above that these parameters are constrained by some theoretical and observational arguments. This issue was studied extensively in
\cite{jm,jac,afjw,cl,ej3,ems,fj,tw,jac2,fos,zfz,ybby,ghlp,omw}. The combination of the
gravitational wave event GW170817 \cite{Abb1} and the gamma-ray burst event GRB 170817A \cite{Abb2}--together with other theoretical and observational constraints--puts a
stringent bound on $c_\sigma$ as (see, e.g., Ref. \cite{omw})
\begin{equation}\label{csig}
|c_\sigma|<10^{-15}.
\end{equation}
On the other hand, imposing that the PPN parameters of  Einstein-Aether theory are identical to those of General Relativity, the stability against linear perturbations in Minkowski background,
vacuum Cherenkov effect and nucleosynthesis constraints imply that (see, e.g., Refs. \cite{fj,omw})
\begin{eqnarray}
&&0\lesssim c_a\lesssim2.5\times10^{-5},\label{ca}\nonumber\\
&&0 \lesssim c_\omega<\frac{c_\sigma}{3(1-c_\sigma)},\label{cw}\nonumber\\
&&0\lesssim c_2\lesssim 0.095,\label{c2}\nonumber\\
&&c_4\lesssim 0.\label{c4}
\end{eqnarray}
Furthermore, in \cite{tzzw}, it was shown that taking $c_\sigma=0$ exactly, the stability of  black holes under  odd-parity  perturbations requires $c_4=0$ along with
$c_1\geq 0$.

Let us discuss the field equation of the theory which we shall need.  The fundamental fields of the gravitational sector are $(g_{\mu\nu},v^\mu,\lambda)$; so, by varying the action (\ref{action}) with respect to $\lambda$, one
obtains the constraint equation (\ref{con}), and varying with respect to the metric $g^{\mu\nu}$ and the aether field $v^\mu$, one respectively obtains
\begin{eqnarray}
&&G_{\mu \nu}+\Lambda g_{\mu\nu}=T^{EA}_{\mu\nu}+8\pi G\,T^M_{\mu\nu}, \label{eqn01}\\
&&\nonumber\\
&&c_{4}\, a^{\alpha}\, \nabla_{\mu}\,
v_{\alpha}+\nabla_{\alpha} \, J^{\alpha}\,_{\mu}+\lambda\,
v_{\mu}=8\pi G\,T^M_\mu,\label{eqn02}
\end{eqnarray}
where
\begin{eqnarray}
&&T^{EA}_{\mu\nu}\equiv\nabla_{\alpha}\, \Big ( J^{\alpha}\,_{(\mu}
\,v_{\nu)}-J_{(\mu}\,^{\alpha}\, v_{\nu)}+J_{(\mu \nu )}\,
v^{\alpha} \Big ) \nonumber\\
&&~~~~~~~~~~+c_{1}\Big (\nabla_{\mu}\, v_{\alpha}\, \nabla_{\nu}\,
v^{\alpha}-\nabla_{\alpha}\, v_{\mu}\, \nabla^{\alpha}\,
v_{\nu} \Big) \nonumber \\
&&~~~~~~~~~~+c_{4}\, a_{\mu}\, a_{\nu}+\lambda\, v_{\mu}\,
v_{\nu}-{1 \over 2} L\, g_{\mu \nu},\label{TEA}\\
&&T^M_{\mu\nu}\equiv-\frac{2}{\sqrt{-g}}\frac{\delta I_M}{\delta g^{\mu\nu}}, \hskip 0.5 cm T^M_{\mu}\equiv-\frac{1}{\sqrt{-g}}\frac{\delta I_M}{\delta v^{\mu}}
,\nonumber
\end{eqnarray}
with
\begin{eqnarray}\label{JL}
J^{\mu}\,_{\nu}=K^{\mu \alpha}\,_{\nu \beta}\, \nabla_{\alpha}\,
v^{\beta}, \hskip 0.3 cm L=J^{\mu}\,_{\nu} \nabla_{\mu}\, v^{\nu}.
\end{eqnarray}
Note that in getting (\ref{eqn01}), we eliminated the term related to the constraint (\ref{con}). Multiplying the aether equation (\ref{eqn02}) by $v^\mu$, one can also derive the Lagrange multiplier
\begin{equation}\label{}
\lambda=c_{4}a^2+v^\mu\,\nabla_{\alpha}\, J^{\alpha}\,_{\mu}-8\pi G\,T^M_\mu v^\mu.
\end{equation}
This section was a necessary recapitulation of the Einsten-Aether Theory theory in its full generality and the pertaining constraints for its viability. Next, we introduce the minimal version of it.

\section{Reduced Einstein-Aether Theory}

Now we shall introduce \textit{Reduced Einstein-Aether Theory} as a nontrivial, theoretically and observationally consistent limit of  Einsten-Aether Theory. The following intuitive argument justifies the motivation for introducing this theory.

As we stated previously, the combination of the gravitational wave observation GW170817 \cite{Abb1} and the gamma-ray burst observation GRB 170817A \cite{Abb2} immediately puts the stringent bound (\ref{csig}) on the coupling constant $c_\sigma$. Since the upper value for $c_\sigma$ is extremely small, this bound suggests that one may strictly take
\begin{equation}
c_\sigma=0,
\end{equation}
which, in turn, requires to strictly take
\begin{equation}
c_4=0,
\end{equation}
as was pointed out in \cite{tzzw} by the theoretical arguments on the stability of black holes under odd-parity perturbations. On the other hand, since the upper bound on the parameter $c_\omega$ in (\ref{c4}) is directly proportional to $c_\sigma$, by taking $c_\sigma=0$, one should also strictly take
\begin{equation}
c_\omega=0.
\end{equation}
Then, from the second and third definitions in (\ref{redefc}), it is immediately worked out that one must strictly take
\begin{equation}\label{c13}
c_1=c_3=0,
\end{equation}
which subsequently means, from the first and the last definitions again in (\ref{redefc}), that
\begin{equation}
c_\theta=3c_2,~~ c_a=0,
\end{equation}
valid strictly. After this reasoning, we see that the only constant that remains nonzero--to have a nontrivial limit of  Einstein-Aether theory--is $c_2$. In other words, $c_2$ is the only coupling constant in  Einsten-Aether Theory that is assumed to be nonzero, but small (close to zero), satisfying the third bound in (\ref{c4}).

To sum up, in  Einstein-Aether theory, we can consistently and confidently take
\begin{equation}
c_\sigma=c_\omega=c_a=0, ~~~~ c_\theta\neq 0, \label{const1}
\end{equation}
or, in the original parametrization,
\begin{equation}\label{c1234}
c_1=c_3=c_4=0, ~~~~ c_2\neq 0,
\end{equation}
satisfying all the theoretical and observational bounds in Eqs. (\ref{csig})-(\ref{c4}) discussed in Sec. II.

The corresponding Lagrangians given in (\ref{lag1}) and (\ref{lag}) then respectively become
\begin{eqnarray}
{\cal L}&=&R-2\Lambda-\frac{c_\theta}{3} \theta^2+\tilde{\lambda}\, (v^{\mu}\,
v_{\mu}+1)\label{lag3}\\
&=&R-2\Lambda-c_2 \theta^2+\lambda\, (v^{\mu}\,
v_{\mu}+1),\label{lag4}
\end{eqnarray}
where $\theta=\nabla_\mu v^\mu$ is the expansion of the aether field. With no any other extra assumptions, we call this nontrivial theory \textit{Reduced Einstein-Aether Theory}.

Studying in the original parametrization, with the choices (\ref{c1234}), we immediately have $K^{\mu \alpha}\,_{\nu \beta}=c_{2}\,\delta^{\mu}_{\nu}\,\delta^{\alpha}_{\beta}$ from (\ref{Ktensor}), and so, $J^{\mu}\,_{\nu}=c_{2} \, \theta\, \delta^{\mu}_{\nu}$ and $L=c_2\,\theta^2$ from (\ref{JL}), where $\theta=\nabla_{\alpha}\,u^{\alpha}$. Then, assuming the aether does not couple to the matter fields (i.e.  $T_\mu^M=0$), we obtain the field equations of Reduced Einstein-Aether theory, from (\ref{eqn01}) and (\ref{eqn02}), as
\begin{eqnarray}
&&G_{\mu \nu}+\Lambda g_{\mu\nu}=\left(\lambda+\frac{c_{2}}{2}\theta^2\right)g_{\mu \nu}+\lambda\,v_\mu v_\nu+8\pi G\,T^M_{\mu\nu},~~~~ \label{eqn03}\\
&&\nonumber\\
&&\nabla_{\mu}\theta+\frac{\lambda}{c_2}\,
v_{\mu}=0, ~~~~ (c_2\neq0)\label{eqn04}
\end{eqnarray}
where, in writing (\ref{eqn03}), we have made use of
\begin{equation}
\lambda=c_2v^\mu\nabla_{\mu}\theta,
\end{equation}
obtained by multiplying (\ref{eqn04}) by $v^\mu$ and using $v_\mu v^\mu=-1$.
 Here, we should also observe that, due to the Bianchi identity ($\nabla^\mu G_{\mu\nu}=0$) and the metric compatibility ($\nabla^\mu g_{\mu\nu}=0$), the covariant divergence of the right-hand side of (\ref{eqn03}) must be equal to zero, which gives the following differential constraint on the Lagrange multiplier $\lambda$:
\begin{equation}
\nabla_\mu\lambda+v_\mu v^\nu\nabla_\nu\lambda+\lambda a_\mu=0,
\end{equation}
upon using (\ref{eqn04}). One important conclusion of this equation is that when the Lagrange multiplier $\lambda=const.\neq0$ (note that $\lambda=0$ gives the Minimal Einstein-Aether theory which will be discussed in the next section), the aether congruence must be geodesics since $a_\mu=0$, and in this case, since $\lambda$ is an arbitrary constant, the covariant divergence of the aether equation (\ref{eqn04}) produces
\begin{equation}
\square\theta+\frac{\lambda}{c_2}\theta=0, ~~~~ (c_2\neq0)
\end{equation}
where $\square\equiv\nabla_\mu\nabla^\mu$. This is the Klein-Gordon equation for the expansion of the aether field with the \textquotedblleft mass'' $m\equiv\lambda/c_2$, where $\lambda$ may be assumed to be positive for having a positive mass. On the other hand, when $\lambda$ is constant, we can also write (\ref{eqn03}) as
\begin{equation}
G_{\mu \nu}+(\Lambda-\lambda)g_{\mu\nu}=\frac{c_{2}}{2}\theta^2g_{\mu \nu}+\lambda\,v_\mu v_\nu+8\pi G\,T^M_{\mu\nu},
\end{equation}
which means that the bare cosmological constant $\Lambda$ is shifted--the Lagrange multiplier behaves like a cosmological constant.

One last observation from the generic equation (\ref{eqn03}) is that the right-hand side is of the perfect fluid energy-momentum tensor form; that is, we can write (\ref{eqn03}) as
\begin{equation}
G_{\mu \nu}+\Lambda g_{\mu\nu}=(\rho+p)v_\mu v_\nu+pg_{\mu\nu}+8\pi G\,T^M_{\mu\nu},
\end{equation}
with
\begin{equation}
\rho=-\frac{c_{2}}{2}\theta^2, ~~~~ p=\lambda+\frac{c_{2}}{2}\theta^2,
\end{equation}
which may be interpreted as the \textquotedblleft energy density'' and the \textquotedblleft pressure'' related to the aether field present in spacetime.

\section{Minimal Einstein-Aether Theory}

As we stated previously, Minimal Einsten-Aether Theory is obtained by letting $\lambda=0$ or, equivalently, $\theta=const.$ (possibly zero) in (\ref{eqn03}). In this case, the filed equations simply reduce to
\begin{equation}
G_{\mu \nu}+\Lambda g_{\mu\nu}=\frac{c_2}{2}\theta^2 g_{\mu \nu} + 8\pi GT^M_{\mu\nu},
\end{equation}
and the aether equation (\ref{eqn02}) is satisfied {\it identically}. At this point, it is worth noting that the so-called Minimal Einstein-Aether Theory studied in \cite{franz} is just a special case of our discussion when $\theta=0$ specifically. In addition to the trivial Einstein limit $c_{1}=c_{2}=c_{3}=c_{4}=0$, we now have a nontrivial limit
$c_{1}=c_{3}=c_{4}=0$ and $\theta=const.$ which is equivalent to the Einstein theory with a shifted cosmological constant $(\Lambda-\frac{c_2}{2}\theta^2)$.  Observe that the contribution from the aether field to the bare cosmological constant is negative as $c_2$ is positive; hence the aether field reduces the cosmological constant. This result is a generalization of \cite{franz} where the expansion is assumed to be zero. In our case, the Kerr metric with a cosmological constant, the Carter metric \cite{cart}, is also an exact solution of the Einstein-Aether theory. The $pp$-wave metric, the AdS wave metric, and all Kerr-Schild-Kundt metrics \cite{ggst}-\cite{gst4} are solutions to this minimal theory.

The unit timelike vector $v^\mu$ satisfies the differential equation $\nabla_{\mu}v^{\mu}=\theta=$ constant. For stationary spacetimes, this equation reduces to
\begin{equation}
\partial_{i}\,( \sqrt{-g}\,v^{i})=\sqrt{-g}\, \theta, \label{eter}
\end{equation}
where the sum above runs over the spatial indices. The zeroth component of the vector is found from $v_{\mu}v^{\nu}=-1$. The only field equation to be solved in Einstein-Aether Theory is
given in (\ref{eter}) which is a linear partial differential equation for the space components of the vector $v^\mu$. For nonstationary cases such as all
Kerr-Schild-Kundt family of metrics we have a similar equation
\begin{equation}
\partial_{a}\,( \sqrt{-g}\,v^{a})=\sqrt{-g}\, \theta, \label{eter1}
\end{equation}
where the sum above covers all coordinates except for a cyclic coordinate on which the metric does not depend. The remaining component of the timelike vector $v^\mu$ is found from $v_{\mu}v^{\nu}=-1$. If the spacetime has no Killing vectors, then one solves the differential equation
\begin{equation}
\partial_{\alpha}\,( \sqrt{-g}\,v^{\alpha})=\sqrt{-g}\, \theta, \label{eter2}
\end{equation}
where the sum above covers all indices. One solves the above differential equation by eliminating one of the components of $v^\mu$ by using $v_{\mu}v^{\nu}=-1$.

\section{Conclusions}
We first considered the limit $c_{1}=c_{3}=c_{4}=0$ in general and reduced the Einstein-Aether field equations to Einstein field equations with a perfect fluid distribution where the expansion parameter is nonconstant.
We then showed that Einstein-Aether theory has a nontrivial limit to Einstein's theory where the aether field has a constant expansion with $c_{1}=c_{3}=c_{4}=0$. This result is equivalent to stating that in this limit, the cosmological constant is shifted and the effective cosmological constant is less than the bare cosmological constant that appears in the Lagrangian or the field equations.   Therefore the aether field in this setting provides a natural way to reduce the cosmological constant which theoretically can be very large yet experimentally very small. From the observational constraints
(\ref{c4}) among the coupling constants, $c_{\theta}=3 c_{2}$ can be considered to be the largest one. Hence another way of interpreting Minimal Einsten-Aether Theory
is that it is an approximation of  Einstein-Aether theory by ignoring the contributions of the other coupling constants except $c_2$ and assuming $\theta=const.$

\end{document}